\documentclass[fleqn,10pt]{wlscirep}

\title{Magnetic ground state of SrRuO$_3$ thin film and
	applicability of standard first-principles approximations to metallic magnetism}

\author[1]{Siheon Ryee}
\author[1,2,*]{Myung Joon Han}
\affil[1]{Department of Physics, Korea Advanced Institute of Science and Technology (KAIST), Daejeon, 305-701, Korea}
\affil[2]{KAIST Institute for the NanoCentury, KAIST, Daejeon, 305-701, Korea}

\affil[*]{mj.han@kaist.ac.kr}

\begin{abstract}
A systematic first-principles study
has been performed to understand 
the magnetism of thin film SrRuO$_3$ which
lots of research efforts have been devoted to but 
no clear consensus has been reached
about its ground state properties.
The relative $t_{2g}$ level difference, lattice distortion
as well as the layer thickness play together in determining the spin order.
In particular, it is important to understand the difference between
two standard approximations, namely LDA and 
GGA, in describing this
metallic magnetism.
Landau free energy analysis and the magnetization-energy-ratio plot clearly show the different
tendency of favoring the magnetic moment formation, and
it is magnified when applied to the
thin film limit where the experimental information
is severely limited. As a result,
LDA gives a qualitatively
different prediction from GGA in the experimentally
relevant region of strain whereas both approximations
give reasonable results for the bulk phase.
We discuss the origin of this difference and 
the applicability of standard methods to the 
correlated oxide and the metallic magnetic systems.
\end{abstract}

\usepackage{graphicx} 

\begin{document}

\flushbottom
\maketitle

\thispagestyle{empty}

\section*{Introduction} 

The study of transition-metal oxide heterostructures and thin films has been
an exciting field of research 
\cite{Ohtomo_2002, Rabe, Mannhart, Chakhalian_RMP, Stemmer}. 
As the controlled synthesis 
and the combination of multiple `mother compounds' become feasible, 
new
possibilities 
can be realized to create the unique functionality and another
angle to look at a long-standing problem from a different perspective.
Simultaneously, however, it requires 
the further technological developments;
not only for synthesizing but also 
for characterizing the materials. For example, 
the site dislocation, interface sharpness, 
stoichiometry and
the oxygen vacancy have often been issues to understand the 
observed phenomena and the discrepancies between
different measurements \cite{Mannhart,Chakhalian_RMP,Stemmer}.

Partly in this regard, first-principles approach has been playing 
a central role in this field from the early stage \cite{Mannhart,Chakhalian_RMP}.
Due to its capability of independent simulation, 
first-principles calculation not just provides the better understanding
of experiments,
but is also useful to predict or suggest
a new system to have desired properties 
\cite{Chakhalian_RMP,Fennie,JHLee,Khaliullin,Hansmann,MJH_nickelate}.
As for the computation methods, however,
there are several limitations and challenges 
to be overcome especially when the 
`(on-site) electronic correlation' becomes important 
\cite{Anisimov_review,Kotliar_RMP}.

In this paper we study the magnetism of SrRuO$_3$ (SRO113) thin film. First of all,
it is an interesting system from both fundamental science
and application point of view \cite{Koster,Thompson,Toyota_SRO,Chang_SRO,Xia_SRO,Ishigami_SRO,Georges_Hund,Rondinelli_SRO,Mahadevan,Gupta_SRO,SRO_half-metal,Si_PRB} .
Its possible applications, for example, to the
bottom electrode and field-effect devices 
have been discussed \cite{Koster}.
Further, a fundamental understanding of its magnetic properties 
is still controversial \cite{Georges_Hund,Toyota_SRO,Chang_SRO,Xia_SRO,Ishigami_SRO,Rondinelli_SRO,Mahadevan,Gupta_SRO,SRO_half-metal,Si_PRB}.
While the ferromagnetism of SRO113 has traditionally been 
discussed from Stoner's picture \cite{Mazin_Ru-based,Chang_SRO,Koster},
recent studies provide new insights
\cite{Georges_Hund,Millis_stoner,Jeong_SRO113}. 
The relationship to its cousins
({\it e.g.}, CaRuO$_3$, Sr$_2$RuO$_4$,
Ca$_2$RuO$_4$, {\it etc}), which
show fairly different characteristics
including antiferromagnetic insulating (AF-I) phase
and superconductivity \cite{Maeno_RMP,Carlo}, should also be clarified.
Thin film SRO113, in particular, 
exhibits the intriguing
thickness-dependent magnetic and metal-insulator transition \cite{Toyota_SRO,Chang_SRO,Xia_SRO,Ishigami_SRO}.
The nature of this transition is far from clear. 
For example, there is an obvious discrepancy in the previous
reports regarding the critical thickness, and
the ground state magnetism has not been clearly 
identified 
\cite{Toyota_SRO,Chang_SRO,Xia_SRO,Ishigami_SRO,Rondinelli_SRO,Mahadevan,Gupta_SRO,SRO_half-metal,Si_PRB}.

Here we perform a systematic and comparative investigation 
of SRO113 based on the standard first-principles 
computation methods.
We elucidate the controversies
in the previous calculations \cite{Chang_SRO,Rondinelli_SRO,Mahadevan,Gupta_SRO,SRO_half-metal,Si_PRB} and partly resolve the unsettled issues
including the thickness and strain dependent transition
of magnetic and electronic properties
\cite{Toyota_SRO,Chang_SRO,Xia_SRO,Ishigami_SRO,Rondinelli_SRO,Mahadevan,Gupta_SRO,SRO_half-metal,Si_PRB}. 
The role of strain, correlation and lattice distortions are
clarified.
We further 
analyze the applicability of primary approximations, namely,
LDA (local density approximation) and GGA (generalized gradient approximation),
to metallic magnetism. Landau free energy analysis
and the magnetization-energy-ratio plot
clearly show the different
tendency of favoring the magnetic moment formation.
In case of thin film, as a result,
LDA gives a qualitatively
different prediction from GGA in the experimentally
relevant region of strain whereas both approximations
give reasonable results for the bulk phase.
We discuss the origin of this difference and 
the applicability of standard methods to the 
correlated oxide and the metallic magnetic systems.
Our work carries
important information for future study
especially when the experimental information
is severely limited just as in the
thin film or heterostructure.

\section*{Results}

\subsection*{Bulk} 
SRO113 is a ferromagnetic metal (FM-M) with Curie temperature of 160~K and
magnetic moment ($M$) of $1.1 - 1.7$~$\mu_{\rm B}$/f.u. \cite{Koster}.
Orthorhombic crystal structure is stabilized below 850 K with
GdFeO$_3$-type distortion \cite{Koster}. To investigate SRO113, we
first optimized the lattice parameters and the result is in good
agreement with the previous studies (TABLE~\ref{lattice})
\cite{Rondinelli_SRO,Zayak_06,Zayak_08,Mahadevan}. A well-known trend
of the underestimated/overestimated lattice parameters by LDA/GGA is
clearly seen. The deviation from experimental lattice values is within
$\pm1.2$ \%.
Rotation and tilting angles 
calculated by LDA and GGA are slightly overestimated 
(within $\sim$1.6--2.2$^\circ$) in comparison with 
the experimental value of
8.61$^\circ$ and 8.60~$^\circ$, respectively.

Ferromagnetic ground state is well reproduced by both 
LDA and GGA with the optimized and experimental lattice parameters.
The calculated moments are summarized in TABLE~\ref{lattice} 
being consistent with the
previous studies \cite{Rondinelli_SRO,Zayak_06,Zayak_08}. About 70 \% of total
moment resides at Ru-site and the other amount comes from O-site due
to the hybridization \cite{Rondinelli_SRO,Mazin_Ru-based}. GGA moment
is larger than LDA by about factor of two. 
Here we note this unusually large difference of
the calculated moments by two different exchange-correlation (XC)
functionals. It may indicate either that GGA overestimates the
ferromagnetism or that LDA underestimates it. This difference
becomes critical when one tries to predict the ferromagnetic
instability in the thin films for which the experimental detections
are largely limited. This point will be highlighted and
further discussed later in this paper.

The GGA trend of stronger (than LDA) preference 
for ferromagnetic solutions is
also found in the magnetic stabilization energy: $\Delta E_{\rm LDA}$
= $\Delta E^{\rm FM} - \Delta E^{\rm PM}$ = $-7.8$ meV/f.u. and
$\Delta E_{\rm GGA}$ = $-106$ meV/f.u. for o-SRO113 (orthorhombic SRO113).  For
c-SRO113 (cubic SRO113), $\Delta E_{\rm LDA}$ = $-14$ meV/f.u. and $\Delta E_{\rm
	GGA}$ = $-102$ meV/f.u..

\begin{figure}[ht]
	\centering
	\includegraphics[width=0.6\linewidth, angle=0]{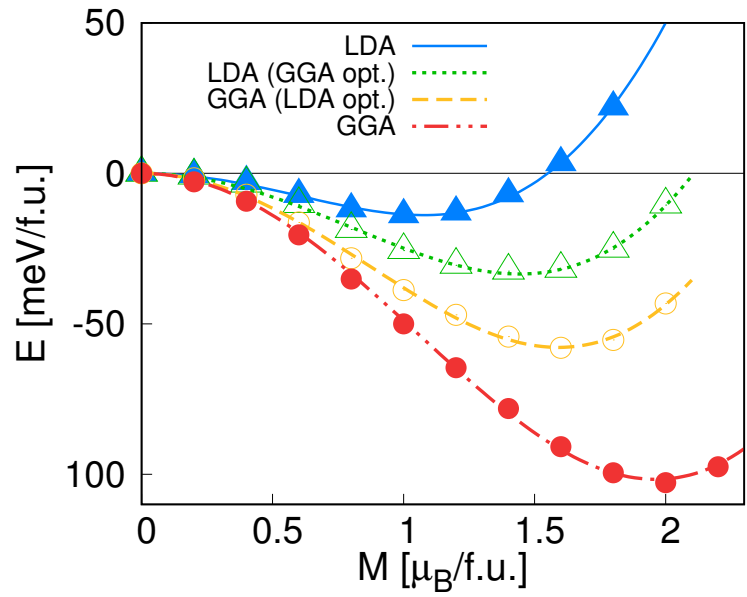}
	\caption{FSM calculation result of total energy for c-SRO113 as a function
		of $M$. The lines are obtained from the fitting to the Landau free energy. The
		open triangles (green) and open circles (yellow) refer to
		the LDA results with the GGA-optimized structure
		and the GGA results with the LDA-optimized structure,
		respectively. The nonmagnetic state is set to be the zero
		reference energy.}
	\label{fsm}
\end{figure}

To have further understanding, we performed the fixed spin moment
(FSM) calculations which have not been reported before 
in these materials. Our
result is summarized in Fig.~\ref{fsm}. The aforementioned magnetic
behavior, namely the enhanced (suppressed) ferromagnetism by GGA (LDA),
is again clearly manifested in this plot (see blue and red lines in
Fig.~\ref{fsm}). Note that this is not just the effect from the different
lattice constants given by GGA and LDA optimization. Even when the
same crystal structures were used, GGA total energy more favors the
ferromagnetic solution than LDA; compare the blue-solid line (filled triangles) 
with the yellow-dashed line (open circles), and compare the green-dotted 
line (open triangles) with the red-dashed-dotted lines (filled circles). 
It is also noted that the GGA-optimized crystal structure 
is more favorable to ferromagnetism than
the LDA structure; compare the blue solid line (filled triangles) 
with the green dotted lines (open triangles), and compare the yellow-dashed
line (open circles) with the red-dashed-dotted lines (filled circles).  

The ferromagnetism in ruthenates has been understood traditionally
based on the
Stoner model \cite{Mazin_Ru-based,Koster,Chang_SRO}.
Therefore, it would be fruitful to estimate the Stoner parameter ($I$)
from our FSM calculation results. Landau free energy can be
written as,
\begin{eqnarray}
{\rm E}(M) = E(0) + (a_2/2)M^2 + (a_4/4)M^4 + \dots,
\end{eqnarray}
where E($M$) is the total energy as a function of magnetic moment (Fig.~\ref{fsm}). The
uniform spin susceptibility is given by
\begin{eqnarray}
\chi^{-1} = \frac{\partial^2 {\rm E}}{\partial {M}^2}{\bigg|}_{M = 0}
= a_2 = \frac{1}{2}\bigg(\frac{1}{{\rm N}({\rm E_F})} - I\bigg),
\end{eqnarray}
where N(E$_{\rm F}$) is the non-spin-polarized density of states (DOS) per spin at Fermi level
(E$_{\rm F}$). The calculated Stoner $I$ and $I$N(E$_{\rm F}$) are
presented in TABLE~\ref{lattice}. The LDA result is in good agreement
with the previous study 
\cite{Mazin_Ru-based}. Both LDA and GGA results
satisfy the Stoner's criterion of ferromagnetism 
and the calculated magnetic moments are in
reasonable range consistent with experiments.
The calculated $I$ and $I$N(E$_{\rm F}$) values by
GGA are larger than those of LDA by $\sim 17$ \% and $\sim 40$ \%, respectively.  Namely,
the stronger preference for FM solution by GGA (than LDA) can also be
seen in these parameters. We once again emphasize that this cannot be
attributed to the lattice effect. It is rather directly related
to the parameterization of XC functional itself and thus
requires further understanding at the more fundamental level. Our
result demonstrates that the special care needs to be paid
in ruthenates and related systems
when one uses first-principles methods even within the standard
XC approximations.

\subsection*{Undistorted thin films: $1a\times 1a$ lateral unitcell}

Now let us turn our attention to the thin film. The previous calculations and experimental results are 
not quite consistent with one another regarding, for example,
the thickness-dependent magnetic states 
on SrTiO$_3$ substrate
\cite{Rondinelli_SRO,Mahadevan,Gupta_SRO,SRO_half-metal,Si_PRB,Toyota_SRO,Xia_SRO,Chang_SRO,Ishigami_SRO,Koster}.
Since SRO113 thin film and its magnetic property have
been studied often within 1$a$ $\times$ 1$a$ lateral unitcell (uc)
\cite{Rondinelli_SRO,Chang_SRO},
we focus on 1$a$ $\times$ 1$a$ films
before considering the effect of tilting/rotational
distortions of RuO$_6$ octahedra.

\begin{figure*}[ht]
	\centering
	\includegraphics[width=\linewidth, angle=0]{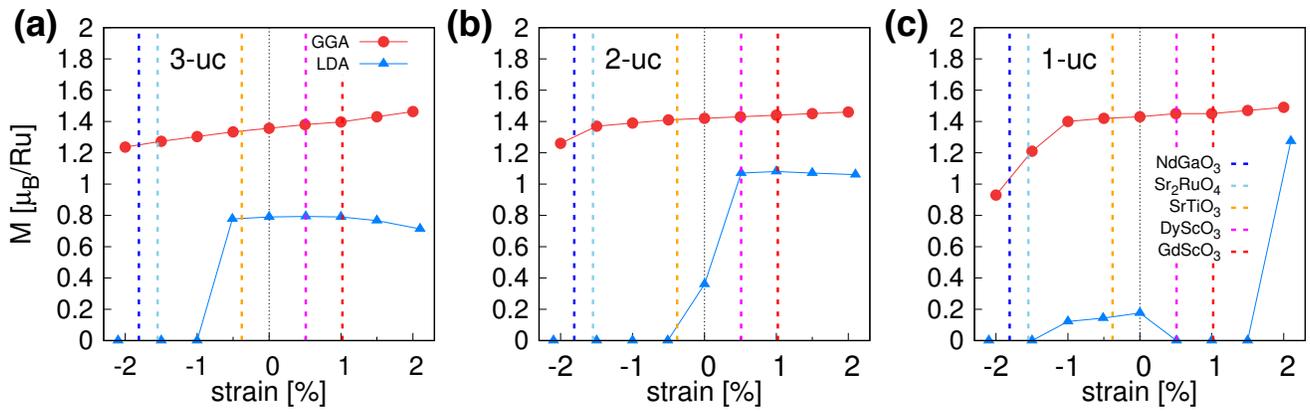}
	\caption{The calculated Ru moments of
		(a) 3-uc, (b)
		2-uc, and (c) 1-uc thick c-SRO113 by LDA (blue solid lines; triangles)
		and GGA (red solid lines; circles) as a function of strain.
		The zero strain is set to the optimized lattice parameters of the bulk
		structure by each XC functional (TABLE~\ref{lattice}). 
		The vertical dashed lines represent
		the strains realizable by NdGaO$_3$ (blue), Sr$_2$RuO$_4$ (light blue),
		SrTiO$_3$ (orange), DyScO$_3$ (magenta), and GdScO$_3$ (red) as
		substrates.}
	\label{layer}
\end{figure*}

Our results are summarized in Fig.~\ref{layer}.
The calculated magnetic moments by GGA (red circles) and LDA
(blue triangles) are presented as a function of strain. 
Fig.~\ref{layer}(a), (b) and (c) shows the result of the  
3-uc,  2-uc, and 1-uc thick SRO113 films, respectively. 
In the 1$a$ $\times$ 1$a$ lateral uc,
antiferromagnetic spin order can not be 
simulated and the finite moment indicates the ferromagnetic order
as in the previous studies \cite{Rondinelli_SRO,Chang_SRO}.
It is noted  that GGA more prefers the
ferromagnetic solution than LDA, which is the same tendency
as observed in the calculations of bulk SRO113.

One general feature found in Fig.~\ref{layer} is that
the ferromagnetism is suppressed
by compressive strain. Both LDA and GGA predict
the zero moment or paramagnetic phase under the large enough compressive strain.
Importantly, however, the difference between GGA and LDA leads to a 
different prediction for the existence of ferromagnetic order
and the critical value of strain below (above) which the magnetic moment 
or ferromagnetic order vanishes (sets in).
In GGA, the ferromagnetic order is robust 
for 3-uc, 2-uc, and 1-uc films, which is significantly different from
the LDA predictions of $\leq -0.5$ (3-uc) -- 0.0 \% (2-uc).

Note that, in the experimentally accessible strain range,
GGA gives a qualitatively different prediction from LDA 
especially for the case of 1-uc.
Vertical dashed lines indicate the strain values 
corresponding to the available
substrates. Within the $-2$ -- $+1.5$ \% range, LDA gives zero (or very small)
moment  while GGA predicts magnetic phase and
the moment is quite large. It should be noted that the zero strain 
in our calculations is
set to be the LDA/GGA-optimized lattice parameters obtained
by each XC functional and therefore two values at a given strain value
represent the different in-plane lattice parameters; 
larger in GGA case and smaller in LDA.

As an example, let us first consider the mono-layer SRO113
grown on NdGaO$_3$ substrate (dark-blue vertical lines)
\cite{substrate}. Due to the
lattice mismatch, this corresponds to the $-1.81$ \% of 
compressive strain situation. For this case,
GGA predicts the ferromagnetic spin order while LDA solution is
nonmagnetic. The same is true for the case of Sr$_2$RuO$_4$
substrate \cite{Anwar} which corresponds to $-1.55$ \% compressive strain.
Our result therefore raises a question
about the predictive power of standard first-principles methodology.
On the one hand, it shows that the special care needs to be paid in the 
first-principles study of correlated oxide thin films and/or
heterostructures even with the standard XC energy functionals.
On the other, our result calls the further method development
which can better describe the electronic correlations
and the structural properties.
We note that in this kind of systems 
conventional experimental techniques are often less useful.
In this sense, correlated oxide heterostructure and thin film pose
a new challenge to the first-principles methodology (see Discussion).

Another notable example is SRO113 film on SrTiO$_3$ substrate
($-0.38$ \% compressive strain).
Contrary to the previous experimental reports
\cite{Toyota_SRO,Xia_SRO,Chang_SRO,Ishigami_SRO},
there is no clear magnetic-to-nonmagnetic and metal-to-insulator (MI) transition
in both of our GGA and LDA results.
In 1-uc case, GGA result is ferromagnetic with $\sim$1.4~$\mu_{\rm B}$/Ru
while a significantly smaller magnetic moment of $\sim$0.2~$\mu_{\rm B}$/Ru
is found in LDA (Fig.~\ref{layer}(c)).
It may seem inconsistent not only with experiments but also with
the previous LDA \cite{Chang_SRO,Rondinelli_SRO} calculations
since they report the magnetic
to non-magnetic transition as the SRO layer thickness is reduced
although the critical thickness is still under debate 
\cite{Toyota_SRO,Xia_SRO,Ishigami_SRO,Rondinelli_SRO,Chang_SRO}.
It should be noted however that in the previous LDA \cite{Chang_SRO} studies,
the experimental SrTiO$_3$
lattice parameter ($3.905\rm \AA$) is taken for the
calculations.
According to our optimized lattice parameters, 
this corresponds to $+0.69$ \% tensile strain for the case of LDA
and $-2.08$ \% compressive strain for GGA, both of which are
significantly different from the experimental situation
of $-0.38$ \% compressive strain.
In fact, our results of $+0.69$ \% tensile and $-0.38$ \%
compressive strain are
consistent with the previous calculations \cite{Rondinelli_SRO,Chang_SRO}. 
At the correct strain of SRO113 on SrTiO$_3$ ($-0.38$ \%),
the LDA magnetic moment gradually decreases from 3-uc
to 1-uc which may be understood as a magnetic to non-magnetic
transition observed in experiments.
In GGA, however, ferromagnetism survives down to 1-uc thickness. 
This behavior is double checked by other codes, namely, OpenMX and ecalj.

It is clear that understanding the SRO113 thin film based on 1$a$ $\times$ 1$a$ is quite limited. 
The tilting and rotation distortions can play important roles
in determining the electronic and magnetic property.

\subsection*{Distorted thin films: $\sqrt{2}a\times\sqrt{2}a$ lateral unitcell}

The octahedral distortion (tilting and rotation of RuO$_6$
cage) and the use of enlarged lateral uc 
can lead to a qualitatively different ground state solution.
The 2-uc-thick SRO113 is found to have FM-M ground state
in the strain range of about $-$3 to +2.5 \%.
At $\sim$3 \%, G-type-like ({\it i.e.,} all the nearest-neighbor couplings are
antiferromagnetic) AF-I is stabilized in both LDA and GGA with Ru-site magnetic moment of 0.74~$\mu_{\rm B}$/Ru (LDA) and 1.51~$\mu_{\rm B}$/Ru (GGA).
It is markedly different from the strained bulk SRO113 for which 
ferromagnetic ground state is fairly robust
over a wide range of strain and G-type AF-I state is not stable \cite{Zayak_06,Zayak_08}.

\begin{figure*}[ht]
	\centering
	\includegraphics[width=\linewidth, angle=0]{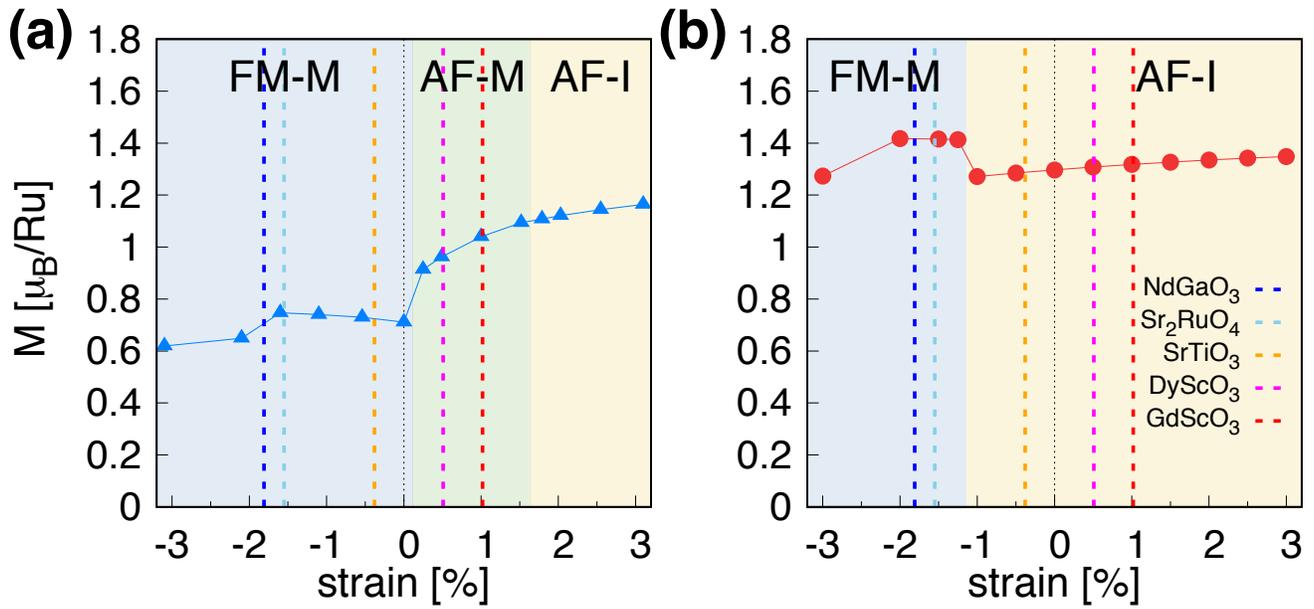}
	\caption{The calculated $M$ of 1-uc $\sqrt{2}$$a$ $\times$ $\sqrt{2}$$a$ o-SRO113
		by (a) LDA and (b) GGA as a function of strain. The different types of ground configurations
		(FM-M, AF-M, and AF-I)
		are presented by background colors. The vertical dashed lines
		represent the strains realizable by NdGaO$_3$ (blue), Sr$_2$RuO$_4$
		(light blue), SrTiO$_3$ (orange), DyScO$_3$ (magenta), and GdScO$_3$
		(red) as substrates.}
	\label{phase}
\end{figure*}

To understand the detailed electronic structure and its relation to the magnetism,
we focus on the thinnest case ({\it i.e.,} 1-uc thick) in the remaining part of this subsection.
The ground state phase diagram as a function of strain is presented in
Fig.~\ref{phase}; (a) LDA and (b) GGA. The first thing to be noted is that
both XC functionals predict FM-M and AF-I ground state in the large compressive
and large tensile strain region, respectively.
The size of magnetic moment is generally larger in GGA than
LDA, which is the same feature found in the previous cases of bulk and undistorted thin film.

In the experimentally more accessible range, however, LDA and GGA
give qualitatively different solutions.
In the GGA phase diagram (Fig.~\ref{phase}(b)), AF-I is the ground state
around zero strain region 
($\leq\pm$1 \% strain)  while LDA predicts either FM-M or antiferromagnetic metal (AF-M).
The AF-M phase of LDA (which is absent in GGA phase diagram) is 
attributed to the smaller energy splitting 
between $d_{xy}$ and $d_{yz,zx}$ states near E$_{\rm F}$.
Thus, for the case of SrTiO$_3$ \cite{substrate},
DyScO$_3$ \cite{substrate} or GdScO$_3$ \cite{Thompson} substrate,
we have two different predictions regarding the ground state property 
from two standard XC functionals.
It therefore raises a serious question about the predictive power of
the current first-principles method to describe the correlated
oxide thin film and/or heterostructure for which the experimental information
is often limited. Further investigation and development are urgently necessary.

\begin{figure}[ht]
	\centering
	\includegraphics[width=\linewidth, angle=0]{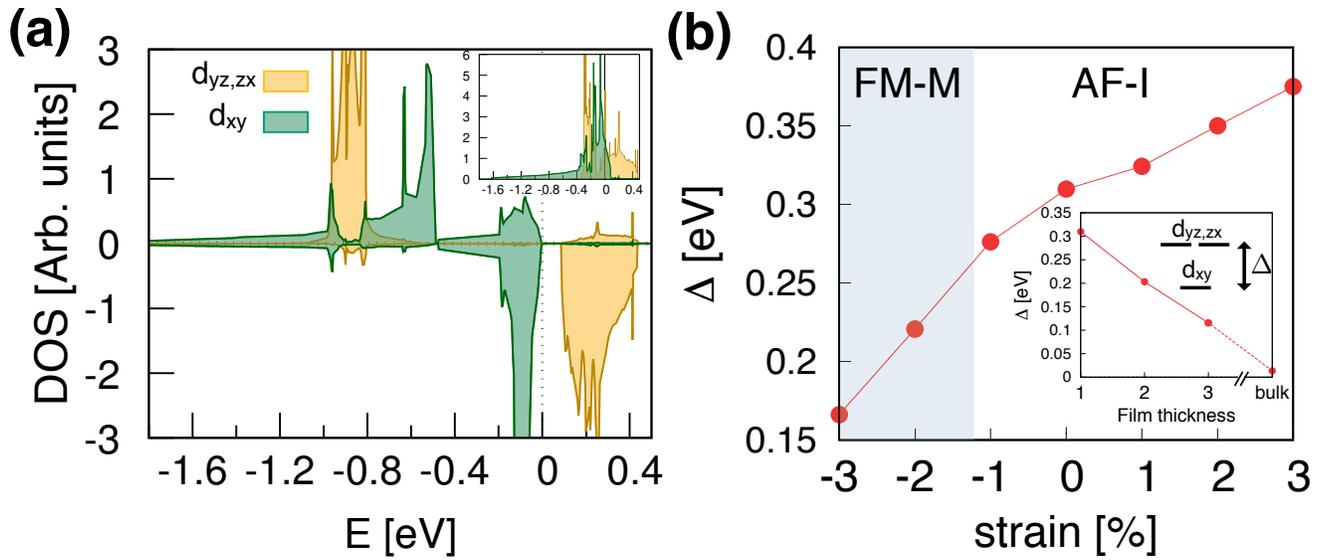}
	\caption{
		(a) The Ru-t$_{2g}$ DOS by GGA for 1-uc o-SRO113
		at $- 0.5$ \% strain. The positive and negative DOS represent the
		up- and down-spin parts, respectively. The Fermi level is set to
		zero. The non-spin-polarized DOS is shown in the inset. 
		(b) The calculated crystal-field splitting, $\Delta$, for 1-uc o-SRO113
		by GGA as a function of strain. The inset shows $\Delta$
		as a function of film thickness (the number of layers) at 0 \% strain.}
	\label{dos}
\end{figure}

Noticeable is our GGA result
of AF-I ground state for SrTiO$_3$ substrate
(see Fig.~\ref{phase}(b)). We emphasize that this 
AF-I phase has never been achieved before by LDA and GGA
\cite{Chang_SRO,Gupta_SRO,Mahadevan,Rondinelli_SRO}.
Since AF-I has previously been obtained 
only by DFT+$U$ \cite{Mahadevan,Gupta_SRO} or DFT+DMFT (DFT + dynamical mean-field theory) \cite{Si_PRB} and it can be consistent
with the recent experiments reporting an insulating state 
carrying no net moment \cite{Ishigami_SRO,Xia_SRO},
it has been discussed that the insulating gap of 1-uc SRO113
is opened by on-site Coulomb correlation \cite{Si_PRB}.
However, Fig.~\ref{dos}(a) clearly shows the band-gap
without the explicit inclusion of $U$ 
while the gap size is smaller than that of DMFT (E$^{\rm DMFT}_{\rm g}$ = $1.0$ eV)
\cite{Si_PRB}. Our result therefore implies that the insulating ground state
of mono-layer SRO113 is not just attributed to the local Coulomb physics within
Ru-$d$ electrons, but the 
homogeneous electron approximation 
can describe the gap opening.
It also shows that the careful
consideration of lattice effect is important to simulate the experimental situation
of complex oxides.


A common trend found in both LDA and GGA phase diagram is that the tensile strain
makes system antiferromagnetic while the compressive strain favors ferromagnetic.
It can be simply understood by defining an energy level difference 
between Ru-$d_{yz,zx}$
and $d_{xy}$ state: 
$\Delta^{\rm LDA/GGA} \equiv \varepsilon^{\rm LDA/GGA}_{yz,zx}
- \varepsilon^{\rm LDA/GGA}_{xy}$.
It is straightforward to compute $\varepsilon$ by using the standard
technique of
maximally localized Wannier function 
\cite{Wannier, Wannier_code}.
At 0 \% strain, $\Delta^{\rm LDA} = 0.33$ and $\Delta^{\rm GGA} = 0.31$ eV.
As the in-plane lattice parameter gets smaller (more compressive strain),
$\Delta$ decreases to 0.15 (LDA) and
0.17 (GGA) eV at $\sim -3$ \% while under tensile strain
it becomes larger to 0.40 (LDA) and 0.38 (GGA) eV at $\sim + 3$ \%.
The same trend is also observed in the 2-uc-thick film; $\Delta = 0.21$ eV (LDA) and 0.20 eV (GGA) at 0 \% strain while $\Delta = 0.27$ eV (LDA) and 0.27 eV (GGA) at $\sim + 3$ \% strain where AF-I is stabilized. The calculated results of $\Delta^{\rm GGA}$ is presented
in Fig.~\ref{dos}(b). In the large $\Delta$ limit (large tensile strain regime),
the separation 
between $\varepsilon_{yz,zx}$ and $\varepsilon_{xy}$ is large and
it approximately corresponds to the (two-band) half-filled case
within the ionic picture of Ru$^{4+}$ (see inset of Fig.~\ref{dos}(a)) as in the case of Ca$_2$RuO$_4$ \cite{CRO}.
Thus the antiferromagnetic spin order is naturally stabilized via superexchange. 
On the other hand,
the small $\Delta$ limit (compressive strain) basically corresponds 
to the bulk SRO113 regime in which ferromagnetic order is stable.

The inset of Fig.~\ref{dos}(b) shows the calculated $\Delta^{\rm GGA}$ as a function
of SRO thickness. A decreasing trend is clearly
noticed as the number of layers increases. Since this is a qualitatively
consistent trend with the magnetic transition
observed in the several experiments as a function of layer thickness,
our result implies that the $\Delta$ plays an important role in the
thickness-dependent
transition. Therefore it is probably not understood solely from N(E$_{\rm F}$)
and Stoner's criterion \cite{Chang_SRO},
but rather comparable with the case of SrVO$_3$/SrTiO$_3$
in which the lifted orbital degeneracy plays an important role
to induce the MI transition \cite{SVO,Held_SVO}.
We do not find any systematic trend in N(E$_{\rm F}$) over the range of strain.
In the sense that the FM-M to AF-I transition 
occurs in between 2- and 1-uc thickness, our GGA result is
not in quite good agreement with experiments which report
the transitions in between 5- and 4-uc \cite{Toyota_SRO}, or
4- and 3-uc \cite{Xia_SRO,Ishigami_SRO}, or 3- to 2-uc
\cite{Chang_SRO}. Further inclusion of electron
correlations might be needed to correctly describe the $\Delta$
and other electronic properties
\cite{Georges_Hund,Held_SVO,correlation1,Tomczak_QSGW,MJH_QSGW,SWJ_QSGW,SR_QSGW}.

It might be informative to see DFT+$U$ \cite{Dudarev} results. In spite of its static Hartree-Fock nature, it has been used to study SRO113 \cite{Gupta_SRO,Mahadevan}. With a fixed $U$ value to a recent cRPA (constrained random phase approximation) result ($U = 3.5$ eV \cite{Si_PRB}) and at a $- 0.5$ \% compressive strain (close to SrTiO$_3$ value), AF-I phase is found to be the ground state. For 1-uc thickness, E$_{\rm g}$ = $1.1$ (1.6) eV and $M = 1.47$ (1.60) $\mu_{\rm B}$/Ru. in LDA+$U$ (GGA+$U$). In 2-uc thickness, E$_{\rm g}$ = $0.9$ (1.3) eV and $M = 1.41$ (1.57) $\mu_{\rm B}$/Ru. in LDA+$U$ (GGA+$U$).

\section*{Discussion} 

Our results show that the prediction of the magnetic ground state of
complex oxide thin film or heterostructure is a challenge to the 
current first-principles methodology. Without further information
such as structural property and/or magnetic order (usually known
from experiment in the case of bulk), it is difficult to make a 
prediction for thin film SRO113.
Note that LDA and GGA are not a bad choice to describe the
bulk SRO113 in the sense that both predict the correct FM-M ground state.
Although some detailed features in the electronic structure are not
well captured by LDA or GGA ({\it e.g.}, Hubbard-like state $\sim -1.3$ eV
below the Fermi energy; see, Ref.~\cite{LHB3}),
these standard approximations provide reasonable descriptions
of this metallic bulk material. Arguably, they are better than other
improved but static methods to take orbital-dependent correlations
into serious account
such as DFT+$U$ \cite{Anisimov_review}. It is therefore surprising to see such a large difference
between LDA and GGA in this system.

We first note that this difficulty is related to the complex nature of correlated oxides
in which the spin, orbital, and lattice degree of freedom are tightly
coupled to each other. On the one hand, it is the reason why the parameter-free
first-principles techniques are strongly requested. On the other hand,
this complexity makes the first-principles description be a greater challenge.
Note that an improved description of electronic correlations cannot immediately be
the right answer unless it can simultaneously give the better description
of the other degrees of freedom such as lattice. For example, if we resort to DMFT
approximation to get a better electronic structure of ruthenates,
we have to describe the force and the lattice relaxation simultaneously
at the same level. This development is technically non-trivial \cite{force1,force2,force3}. Similar issues and related intrinsic ambiguities can also be found in other techniques such as SIC-DFT (self-interaction corrected DFT) and hybrid functional.

Here it is instructive to recall that the metallic magnetism
has been a challenge to the first-principles calculation
\cite{Terakura,Aguayo,Mazin_ironSC,Ortenzi}.
A classical
example is bcc Fe for which minimum energy solution of LDA
is paramagnetic hcp phase (with the optimized
lattice parameter) \cite{Terakura} unlike Co and Ni \cite{TM}.
In other words, without {\it a priori}
knowledge of ferromagnetism, the situation of bcc Fe is similar 
with SRO113 thin film; paramagnetic ground state by LDA 
and ferromagnetic in GGA.
Since the use of experimental lattice parameter
gives the correct ferromagnetic solution in both LDA and GGA,
bcc Fe is an easier case in the practical sense.
In SRO113 thin film, on the other hand, the limitation of
computation method is magnified due to no {\it a priori}
information of structure and magnetic order.

\begin{figure}[ht]
	\centering 
	\includegraphics[width=0.6\linewidth, angle=0]{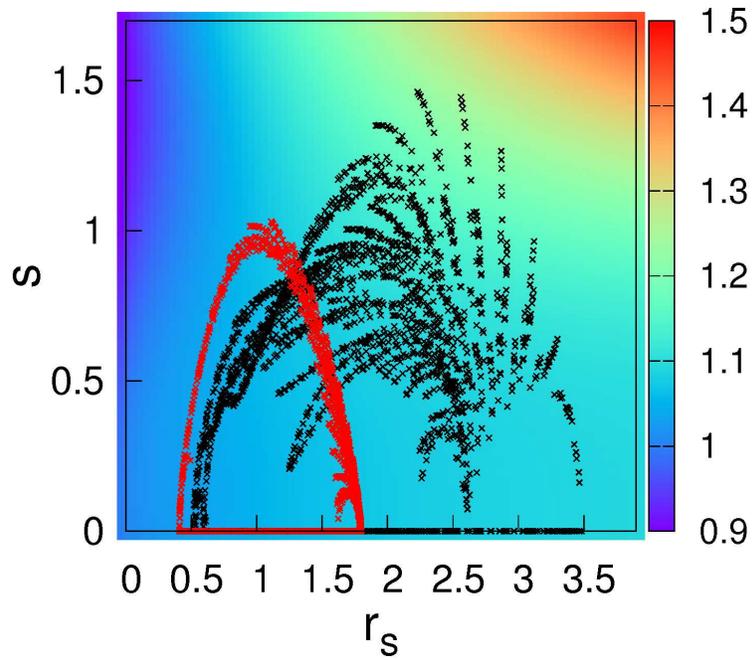}
	\caption{The calculated $\alpha_{\rm XC}$ for c-SRO113 (black crosses)
		and bcc Fe (red crosses). 
		About 85000 real-space grid points are obtained from the self-consistently converged
		electronic structure within LDA. Experimental lattice parameters are used. $\zeta$ is fixed to 0.02 which is the average value of c-SRO113. The $\alpha_{\rm XC}$ change is negligible in the range of $0 < \zeta < 0.5$.}
	\label{map}
\end{figure}

A more detailed nature of enhanced ferromagnetism by GGA
functional can be seen in Fig.~\ref{map} where the PBE (Perdew-Burke-Zunger) correction of spin-polarized energy density
with respect to LDA is presented as a function of inverse 
density ($r_s$) and density gradient ($s$) for the relative spin-polarization ($\zeta$). 
The calculated values of
\begin{eqnarray} 
\alpha_{\rm XC} \equiv {{\Delta}\epsilon^{\rm PBE}_{\rm XC}(r_s, \zeta, s) \over
	{\Delta}\epsilon^{\rm LDA}_{\rm XC}(r_s, \zeta)}
\end{eqnarray}
are represented by color. Here  
${\Delta}\epsilon^{\rm PBE}_{\rm XC}(r_s, \zeta, s) = \epsilon^{\rm unif}_{\rm X}(r_s,0)[F_{\rm XC}(r_s,\zeta,s)-F_{\rm XC}(r_s,0,s)]$ and
${\Delta}\epsilon^{\rm LDA}_{\rm XC}(r_s, \zeta) = f(\zeta) [\epsilon^{\rm unif}_{\rm XC}(r_s,1) - \epsilon^{\rm unif}_{\rm XC}(r_s,0)]$
refer to the spin-polarized part of XC
energy density within PBE and LDA, respectively.
For the definitions of $\epsilon$, $r_s$, $\zeta$, $s$, $F_{\rm XC}(r_s,\zeta,s)$, and $f(\zeta)$, we follow the original PBE \cite{PBE} and LDA-PZ (Perdew-Zunger) \cite{CA-PZ} papers. By definition, 
$\alpha_{\rm XC}$ represents
the energy gain by GGA-PBE over LDA.
In the yellow-red colored region of Fig.~\ref{map},
GGA-PBE prefers the magnetic solution more than
LDA while in the dark blue region LDA-like solution
is favored. About 85000 real-space grid
points (using OpenMX code) are
plotted in this map for bcc Fe (red-colored points)
and c-SRO113 (black-colored points), respectively.
It is clear that
both bcc Fe and c-SRO113 have large portion of grid points
in the region of $\alpha_{\rm XC}$ larger than unity.
It is responsible for the larger spin splitting ($\sim \partial \Delta \epsilon_{\rm XC} / \partial \zeta$) and thus the moment formation enhanced by PBE correction.

Since many interesting aspects of complex oxide films
and interfaces are directly
or indirectly related to the metallic magnetism \cite{MM2, MM1, Li, Bert, Hoffman},
it is strongly requested to develop more reliable 
computation methods.
One promising aspect is the same overall trend given
by LDA and GGA. For example, the moment is suppressed by compressive strain
in the case of $1a\times 1a$ lateral uc
(Fig.~\ref{layer}(a) and (b)). FM-M and AF-I phase in the large compressive
and large tensile strain, respectively, are also consistently reproduced
by LDA and GGA (Fig.~\ref{phase}). Further investigation of
parameterization hopefully gives a way to improve approximations.

\section*{Summary}
From a systematic first-principles investigation,
we elucidate the controversies
in the previous calculations and experiments. 
A part of issues has been clarified including
the thickness- and strain-dependent phase transition as well as
the ground state property of monolayer SRO113.
The role of strain, correlation and lattice distortions are explored in detail.
Applicability of LDA and GGA has been discussed
in the realistic material context by using
Landau free energy analysis and 
magnetization-energy-ratio plot.
While the overall feature is commonly reproduced by both
approximations, they give qualitatively
different predictions for the case of thin film 
in the experimentally relevant region of strain.
Our work provides useful information
not only for ruthenates and other correlated oxides, 
but also for further first-principles
studies in the related systems such as metallic magnetism.

\section*{Methods}

We choose LDA parameterized by Perdew
and Zunger \cite{CA-PZ} and GGA
by Perdew, Burke, and Ernzerhof \cite{PBE} as two representative
XC functionals. The data obtained by these two
functionals is presented as our main results and
discussed in a comparative way. 
We used
projector-augmented-wave (PAW) method \cite{PAW} as implemented in the
VASP software package \cite{VASP}. While VASP-PAW is used to obtain
our main results, we also double checked some cases with other methods,
confirming that our conclusions are valid. The other codes we used
include OpenMX \cite{openmx}, which is based on the norm-conserving
pseudopotential and pseudoatomic orbitals, and ecalj \cite{ecalj} based on
the full-potential `LMTO' or `PMT' basis \cite{PMT}.

Tetrahedron method with Bl{\"o}chl correction has been used for
Brillouin zone integration \cite{tetrahedron}, which we found is
important to achieve the enough accuracy. In some cases the use of
Gaussian broadening is found to give qualitatively different
predictions depending on the broadening value.
For bulk structure optimizations, 
12 $\times$ 12 $\times$ 12 and 8 $\times$ 8 $\times$ 6
\textbf{k}-points were used for 5-atom uc of c-SRO113
and 20-atom uc of o-SRO113, respectively. The
force criterion of 1 meV/{\AA} was adopted for structural
optimization. This choice of \textbf{k} meshes and the 500 eV energy
cutoff were found to be satisfactory from our convergence test. For
the electronic structure calculations, however, we took a denser
\textbf{k}-grid of 16 $\times$ 16 $\times$ 16 for c-SRO113 and 12
$\times$ 12 $\times$ 10 for o-SRO113 to calculate DOS
for the given optimized structures. To simulate SRO113 thin films, we
considered the 3-uc, 2-uc, and 1-uc slab geometries terminated with
SrO layers which is known to be the case in experimental situations
\cite{Koster}. The \textbf{k}-grids used for slab calculations were 12
$\times$ 12 $\times$ 2 for c-SRO113 and 8 $\times$ 8 $\times$ 2 for
o-SRO113. The vacuum thickness is $\geq$15 ${\rm \AA}$ which is large
enough to simulate the experimental situation. The Ru radius is set to 1.402 {\AA} and the Ru moment dependence on this setting is negligible ($\sim$ 0.01~$\mu_{\rm B}$).

Epitaxial strain is one of the key control parameters in the thin film
growth. It is therefore of crucial importance to determine how to
simulate the in-plane lattice parameter corresponding to the
experimental situation. Note that this is a non-trivial issue because
LDA- and GGA-optimized lattice parameters are both in general
different from the experimental values. While the former tends to
underestimate the lattice parameter, the latter overestimates (see
TABLE~\ref{lattice}). Thus, the naive choice of a XC functional or
taking experimental lattice value can easily be misleading. Previous studies
are not quite consistent in this regard. In the current study, we take
the optimized lattice constant with a given XC functional as the
reference value ({\it i.e.}, zero strain) and define the
compressive/tensile strain value with respect to it. This can be the
most reasonable way to simulate the experiments and can serve as a
solid reference for the future study. This point should be kept in
mind when our results are compared with the previous calculation
results \cite{Chang_SRO,Gupta_SRO}.

\bibliography{ref}

\section*{Acknowledgements}

We thank T. Ozaki for useful discussion. We were supported by Basic Science Research
Program through the National Research Foundation of Korea (NRF) funded by the Ministry of Education
(2014R1A1A2057202). The computing resource is supported by
National Institute of Supercomputing and Networking / Korea Institute
of Science and Technology Information with supercomputing resources
including technical support (KSC-2015-C2-011).

\section*{Author contributions}

S.R. performed the calculations under the supervision of M.J.H. Both authors analyzed the results and wrote the paper. 

\section*{Additional information}

The authors declare no competing financial interests.

\newpage

\newpage

\newpage

\newpage

\newpage

\newpage
\begin{table*}[ht]
	\centering
	\begin{tabular}{l l c c c c c c}
		\hline \hline
		System &\ \ \ \  Type &\ \ \ \ $a$ [\AA] &\ \ $b$ [\AA] &\ \ $c$ [\AA] &\ \  Stoner $I$ [eV] &\ \ $I$N(E$_{\rm F}$) &\ \ $M$ [$\mu_{\rm B}$/f.u.]	 \\
		\hline
		&\ \ \ \ Exp.   &\ \ \ 3.92		&\ \ \ 3.92	&\ \ \ 3.92 &\ \ \  --- &\ \ \	--- &\ \ \	---  \\
		c-SRO113 &\ \ \ \ LDA &\ \ \ 3.890 &\ \ \ 3.890	&\ \ \ 3.890 &\ \ 0.44	&\ \ 1.28 &\ \ 1.04 (0.67) \\	&\ \ \ \ GGA  &\ \ \ 3.988 &\ \ \ 3.988	&\ \ \ 3.988 &\ \ 0.51 &\ \ 1.86 &\ \ 2.00 (1.38) \\							
		\hline
		&\ \ \ \  Exp. \cite{SRO113_struct}  &\ \ \ 5.5670 &\ \ 5.5304 &\ \ 7.8446 &\ \ \  --- &\ \ \	--- &\ \ \	--- \\
		o-SRO113        		&\ \ \ \ LDA     &\ \ \ 5.4998 &\ \ 5.4852 &\ \ 7.7503 &\ \ 0.46  &\ \ 1.27  &\ \ 0.82 (0.56) \\
		&\ \ \ \ GGA     &\ \ \ 5.6313 &\ \ 5.6221 &\ \ 7.9500 &\ \ 0.53   	&\ \ 1.75   &\ \ 1.99 (1.40) \\
		
		\hline \hline
	\end{tabular}
	\caption{The calculated lattice constants, Stoner $I$, 
		$I$N(E$_{\rm F}$) and magnetic moments for c-SRO113 and o-SRO113
		by LDA and GGA. For comparison, the experimental values
		are also presented. The experimental lattice constants of c-SRO113 are taken to be the pseudocubic values of o-SRO113. The $M$ values in parentheses represent the
		magnetic moments at Ru sites. }
	\label{lattice}
\end{table*}

\end{document}